\documentclass[11pt]{article}
\usepackage{geometry}                % See geometry.pdf to learn the layout options. There are lots.
\usepackage{graphicx}
\usepackage{amssymb,amsfonts,amsmath}
\usepackage{pifont,topcapt,booktabs}
\usepackage{epstopdf}
\newcommand{\ce}{{\it C. elegans\ }}
\newcommand{\cep}{{\it C. elegans}}
\newcommand{\be}{\begin{eqnarray}}
\newcommand{\ee}{\end{eqnarray}}
\newcommand{\la}{\langle}
\newcommand{\ra}{\rangle}
\title{Information content of colored motifs in complex networks}
\author{Christoph Adami$^{1,2,4,\footnote{Contact author}}$\and Jifeng Qian$^{1}$\and Matthew Rupp$^{3,4}$\and Arend Hintze$^{1,3,4}$\\ \mbox{}\\
\it $^1$Keck Graduate Institute of Applied Life Sciences, 535 Watson Drive, Claremont, CA 91711\\ 
\it $^2$Department of Microbiology and Molecular Genetics\\ 
\it $^3$Computer Science and Engineering\\
\it $^4$BEACON Center for the Study of Evolution in Action\\ 
\it Michigan State University, East Lansing, MI 48824\\
}
\begin{document}
\maketitle
\abstract{We study complex networks in which the nodes of the network are tagged with different colors depending on the functionality of the nodes (colored graphs), using information theory applied to the distribution of motifs in such networks. We find that colored motifs can be viewed as the building blocks of the networks (much more so than the uncolored structural motifs can be) and that the relative frequency with which these motifs appear in the network can be used to define the information content of the network. This information is defined in such a way that a network with random coloration (but keeping the relative number of nodes with different colors the same) has zero color information content. Thus, colored motif information captures the exceptionality of coloring in the motifs that is maintained via selection. We study the motif information content of the \ce brain as well as the evolution of colored motif information in networks that reflect the interaction between instructions in genomes of digital life organisms. While we find that colored motif information appears to capture essential functionality in the \ce brain (where the color assignment of nodes is straightforward) it is not obvious whether the colored motif information content always increases during evolution, as would be expected from a measure that captures network complexity. For a single choice of color assignment of instructions in the digital life form Avida, we find rather that colored motif information content increases or decreases during evolution, depending on how the genomes are organized, and therefore could be an interesting tool to dissect genomic rearrangements.
}
\mbox{}
\vskip 0.5cm

\newpage
\section{Introduction}
One of the most common strategies to understand a complex system is to analyze it in a hierarchical manner. For example, in biology, we attempt to unravel a cell's function by finding all of its parts and understanding how they relate to each other. Because the rules of interactions between cellular components are very complicated, the cell is much more than the sum of its parts: the parts and their interactions form a network, whose properties we can analyze. On the next level, an organism is much more than the sum of its cells, and a society of organisms, in turn, much more than the sum of its members. Thus, networks can be used to study social interactions between individuals also, allowing us to understand the dynamics of groups from the perspective of the mathematics of networks. 

While the ``science of networks"~\cite{AlbertBarabasi2002,Newman2003,Newmanetal2006} has developed tremendously in the last ten years, a comparison of networks across different disciplines, or even of networks within one discipline (such as the protein-protein interaction networks of different organisms) has not really been possible except on the level of the connectivity patterns alone. The complexity of a network--or even perhaps its capacity to perform particular functions--is difficult to quantify, simply because complexity is a multi-faceted concept that as yet does not have an empirical basis. Many different approaches to quantifying complexity exist~\cite{Adami2002b} (a non-exhaustive list is presented in Ref.~\cite{Lloyd2001}), ranging from assessing the complexity of a system's structure~\cite{Lofgren1977,Papentin1980,Papentin1982,ThomasReif1993,Thomasetal2000,SoloveichikWinfree2006,Ahnertetal2010}, or the complexity of the  sequence giving rise to that structure~\cite{Kolmogorov1965,LempelZiv1976,EbelingJimenezMontano1980, LiVitanyi1997,GellMannLloyd1996,AdamiCerf2000}, to quantifying the {\em function} of the sequence or system~\cite{McShea2000,Szostak2003,Hazenetal2007}.  What the measures of complexity have in common is that they all  attempt to capture ``that which increases when self-organizing systems organize themselves"~\cite{Bennett1995}. 

If a network is a succinct description of any complex system, shouldn't a measure of network complexity be the concept that unifies attempts to
attach a number to our intuitive understanding of complication? Unsurprisingly perhaps, a network's complexity appears to be as difficult to quantify as any other complex system. Several attempts exist in the literature~\cite{Wilhelm2003,MeyerOrtmanns2004,Claussen2007,WilhelmHollunder2007}, reviewed in~\cite{KimWilhelm2008}. 

Here we develop and study a measure that attaches a number to a network so that it can be ranked and compared to other networks, and that allows us to track network evolution. This complexity measure is based on the theory of information, and is closely related to a measure that has been proposed to study the complexity of genes. Without a network complexity measure it is not possible to correlate complexity with function. Armed with such a measure, however, we should be able to understand for example how different types of networks react to damage, something that is important for molecular networks as for neural networks, ecological networks, or our cyber infrastructure.

Information is perhaps the central commodity of a technologically advanced society. We use information to order the world around us, make predictions about the world that allow us to function within it, and to encode our knowledge so that it can be passed on to future generations. But while information is an intuitive notion in our day-to-day life, it also has a precise mathematical formulation that meshes perfectly with our intuitive understanding. The theory of information due to Shannon~\cite{Shannon1948} allows us to quantify the amount of information in a book, say, or on a CD or on the hard drive of a computer. It also allows us to study information transmission as well as ways to protect information from noise. Because the theory of information is mathematical in nature, it applies to any information anywhere, in particular to the information stored in our genes. And while this information is not written in the ones and zeros of computers, or the letters of an alphabet, it is written in the language of biochemistry: the nucleotides A, C, G, and T or the twenty amino acids that proteins are made of. 

We have recently described how the information content of genes can be measured from biological sequence information alone~\cite{AdamiCerf2000,Adami2002b,Adami2004}. This information content is measured as the deviation from the expected sequence of a random gene, by recording the frequency with which each symbol appears at any particular position within the sequence. Thus, a highly conserved nucleotide at a particular sequence position indicates strong selection for function there (and thus high information content), while a position where each nucleotide appears with equal probability--the random expectation--stores little or no information about the function of that gene given the particular environment. Because the probability distribution of symbols in the sequence is shaped by Darwinian selection within the environment in which the organism that harbors that sequence lives, it is immediately clear that this information is necessarily {\em functional}, that is, useful to the organism. Qualitatively speaking, this information is used by the organism in order to make predictions about its environment that are better than chance~\cite{Adami2004}. In other words, we expect the information content of genes to correlate with fitness, which has been shown to be the case in at least two different computational systems~\cite{Adamietal2000,Huangetal2004,Ofriaetal2008,HintzeAdami2008} and one biochemical one~\cite{Carothersetal2004}.

A network, if it describes a functioning entity (such as a cell, a brain, the internet, or a group of friends), can be seen as an information-rich structure. Clearly, the nodes and edges carry meaning in such a network, because a rearrangement of the nodes and edges would describe an entirely different system, or at least one with severely impaired function. This meaning, of course, is relative to the environment in which the network functions, just as the meaning of genes is context-dependent. How then can we measure the information that is stored in networks? 
Previous approaches have studied the information contained in degree-degree correlations (the assortativity of the network) to study how functional constraints affect network structure, for undirected~\cite{SoleValverde2004} as well as directed~\cite{Piraveenanetal2010} networks.  Another information-theoretic approach focused on the entropy of randomized ensembles of networks constrained by degree distribution, degree correlation, and community structure~\cite{Bianconi2008}. Here, we take a different approach and instead of considering the degree distribution as the ``degree of freedom" that provides entropy to a network, we study the {\em subnetworks} (sometimes called subgraphs, or motifs)\cite{ShenOrretal2002,Miloetal2002,Riceetal2005} of a network that are obtained when we break up a network into its components, just as we break up a gene into its nucleotide alphabet. 

\begin{figure}[!htbp] %  figure placement: here, top, bottom, or page
   \centering
   \includegraphics[width=4in]{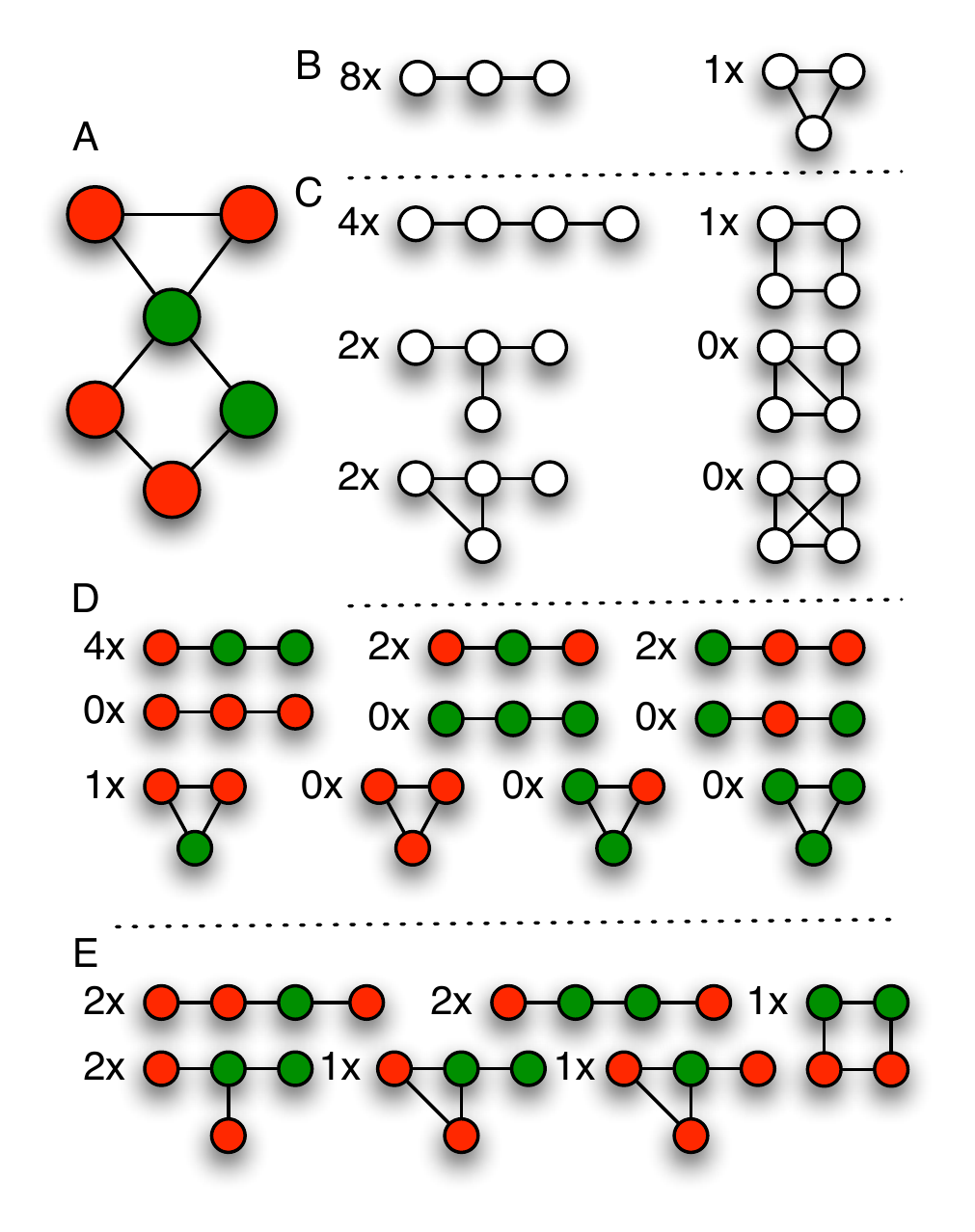} 
   \caption{Motif counting. A: A six-node undirected example graph with two colors can be seen to be made from various motifs. B: For the size-three alphabet (motifs made from three nodes), we find two different structural motifs. C:  For a size-four alphabet we see that two of the six possible structural motifs do not occur in the example graph. D: The colored motif frequencies (three node, two colors) for the network in A. E: Colored motif frequencies in the six-node graph shown in (A) for two colors and four nodes.}
   \label{colmotifs}
\end{figure}

There is some freedom in defining the ``network alphabet": we can use subgraphs of two, three, four nodes, or more. Naturally, subgraphs with more nodes give rise to a network alphabet with more letters (motifs). But once we settle on an alphabet, we can obtain the frequency of each motif in the network, for example as in Fig.~\ref{colmotifs}, where we illustrate the procedure of motif counting for a simple graph of six nodes only.
%Fig. 1
Using the frequency of motifs we can estimate {\em motif probabilities} just as we can estimate the probability to find words in an English sentence using the frequency of words in a text. For the latter case, it is possible to estimate the information content of English text as compared to random sequences of words, an exercise that Shannon already conducted in 1951~\cite{Shannon1951}.  If we assume that a random sequence of words contains no information, then the deviation from the uniform distribution (probability of each word of a given length appearing with equal probability) could be used to distinguish and perhaps classify functional (that is, meaningful) text from gibberish. Indeed, such a test was recently used to distinguish living from non-living matter both in biochemistry and in ALife~\cite{Dornetal2011}. 
In the same vein, it is possible that functional networks differ significantly from random networks in the subgraph utilization, and we can study this difference by estimating the {\em motif information content}. However, it is also not surprising that some of the differences in motif utilization across networks that have been noted previously~\cite{Miloetal2002,Miloetal2004} could be due to constraints imposed by the degree distribution, or other constraints imposed by the growth-process of the network~\cite{HintzeAdami2010}. In other words, it is possible that the non-random ``expression" of structural motifs could be a ``spandrel" of cellular complexity~\cite{SoleValverde2006}.

But in fact, it is not difficult to see that networks contain more information than their topology (that is, the local patterns of connections) alone. Imagine, for example, a network of friends that know each other from high school, say, together with their friends. While we can learn a lot about common interests by looking at the clusters and the type of subgraphs that occur often, this approach assumes that all the nodes (and all the edges, for that matter) are qualitatively the same. However, more information can be gleaned from the network if we attach tags to each node or edge to classify the nodes or edges. For example, we can assign the tags ÒmaleÓ and ÒfemaleÓ to each node, or we can assign the tag Òhigh-schoolÓ and Òafter-high-schoolÓ to the edges that define the relationship between the nodes (referring to the time the two nodes became friends). If we color the graph according to these tags, the subgraph alphabet suddenly becomes much larger because each motif now comes in a variety of colorations. Here, we limit ourselves to colored nodes (leaving edges uncolored), and define an alphabet of colored motifs that we can use to calculate network information content. We show an example of colored motif counting for a six-node graph in Figure~\ref{colmotifs}C and D, using only two colors.

\section{Motif entropy and information}
Entropy in information theory~\cite{CoverThomas1991} is a measure of the uncertainty about the identity of objects in an ensemble. Let $X$ be a random variable describing the structural (or topological) motifs of a network, given the size of motifs (different motif sizes define a different set of possible topological motifs). $X$ can then take on the states  $x_1, ..., x_N$, where $N$ is the number of possible motifs of the given kind. Note that even when the number of nodes is fixed, the number of possible topological motifs still depends on the kind of edges that are allowed in the network (directed or undirected), and whether ``self-edges" are allowed.  
If $q_i$ are the probabilities to find motifs $x_i$, we can define the topological motif entropy as
\be
H_{\rm top}(X)=-\sum_{i=1}^N q_i\log_2 q_i\;.\label{topent}
\ee

Each network has a particular topological motif entropy $H_{\rm top}(X)$ that reflects the motif ``utilization". We can determine whether the distribution of motifs in a network is functionally constrained, by randomizing the edges in the network (while keeping the edge distribution, for example, unchanged). Each randomization will create an instance $H_{\rm top}^R(X)$. The topological information content of the network would then be
\be
I_{\rm top}(X)=\la H_{\rm top}^R(X)\ra-H_{\rm top}(X)
\ee
with $H_{\rm top}(X)$ from Eq.~(\ref{topent}), and where $\la H_{\rm top}^R(X)\ra$ is the topological motif entropy averaged over different edge-randomizations of the network. This definition is formally the equivalent of the definition of the information content at a single nucleotide or residue site $X$~\cite{Adami2004}. 

If nodes can carry colors, they add an element of uncertainty even if the structure of the motif is given, because
each particular topological motif can, given the possible colors that nodes can take on, appear in different colorations. Many of these colorations may be meaningless or downright detrimental for a functioning organism. 
We can quantify the functional constraints that affect colored motifs by studying the {\em color entropy} of a particular structural (topological) motif.  If a particular structural motif $x_i$ is now interpreted as a random variable $Y_i$ that can take on the states $y^{(i)}_1, ..., y^{(i)}_M$ (its possible colorations) with probabilities $p^{(i)}_1,...,p^{(i)}_M$, we can define the {\em color entropy} $H_{\rm color}(Y_i)$ of this motif by measuring how many times each of the colorations $y^{(i)}_j$ appears in the network:
\be
H_{\rm color}(Y_i)=-\sum_{j=1}^M p^{(i)}_j\log_2 p^{(i)}_j\;. \label{colent}
\ee

The average color entropy of motifs in the network is then
\be
H_{\rm color}=\sum_i q_i H_{\rm color}(Y_i)\;. \label{colentav}
\ee
The total entropy of motifs, obtained by counting all possible colored motifs (within each class of motif sizes) is simply given by the sum of the topological and color entropy by virtue of the grouping axiom of information theory~\cite{Ash1965}, i.e., 
\be 
H_{\rm total}=H_{\rm color}+H_{\rm top}\;. \label{tot_ent}
\ee
However, this decomposition does not allow us to determine whether more information is stored in the topology or the functional assignment of nodes, because the baseline (unselected) distribution of motifs depends strongly on the method of randomization used. Furthermore, given a color assignment, an edge randomization automatically implies a color randomization, that is, color information and topological information cannot strictly be separated. 

Nevertheless, we can calculate the information content of motif coloration by randomizing the colors in each network, while keeping the relative numbers of colors unchanged. In this way, we introduce $\la H_{\rm color}^R\ra$, which is calculated just as Eq.~(\ref{colent}) but using a color-randomized version of the network, and averaged over a sufficient number of such randomizations. The color information content
of the entire network is then simply
\be
I_{\rm color}=\la H_{\rm color}^R\ra-H_{\rm color}\;. \label{colinfo}
\ee

\subsection{Motifs in the {\it C. elegans} brain}
To test these measures, we can analyze motifs in the network of synaptic and gap-junction connections of the neuronal network of the nematode \cep.  This network controls one of the most well-understood complex biological systems to date, and  most of the network architecture of the 302 neurons of the hermaphrodite worm is known from experimental work~\cite{Whiteetal1986,HallRussell1991} as well as recent reconstructions~\cite{Varshneyetal2009}. The most up-to-date wiring information covers 279 neurons of the somatic nervous system, excluding 20 neurons of the pharyngeal system and three neurons that appear to be unconnected from the rest~\cite{Varshneyetal2009}. There are 3,606 edges between these nodes, of which some (the synaptic connections) are directed, while gap-junctions are undirected. In our analysis of this network, we describe an undirected edge as a ``bi-directional" edge,  and also place bi-directional edges between nodes if synaptic connections run in both directions between the nodes.

\subsubsection{Two-node colored motifs} In previous work that analyzed structural motifs only~\cite{SpornsKoetter2004,Reigletal2004,Songetal2005}, the uni-directional two-node motif was found to be unremarkable (in the sense that the probability with which it was observed in the actual \ce network was not significantly different from the frequency observed in an edge-randomized version), while the bi-directional motif was deemed over-represented~\cite{Reigletal2004,Songetal2005}. We can look at both of those motifs in terms of the exceptionality of their colorations, by coloring neurons according to three possible functional tags, such as motorneuron (blue), sensor neuron (green), or interneuron (red). 

\begin{figure}[!htbp] %  figure placement: here, top, bottom, or page
   \centering
   \includegraphics[width=5in]{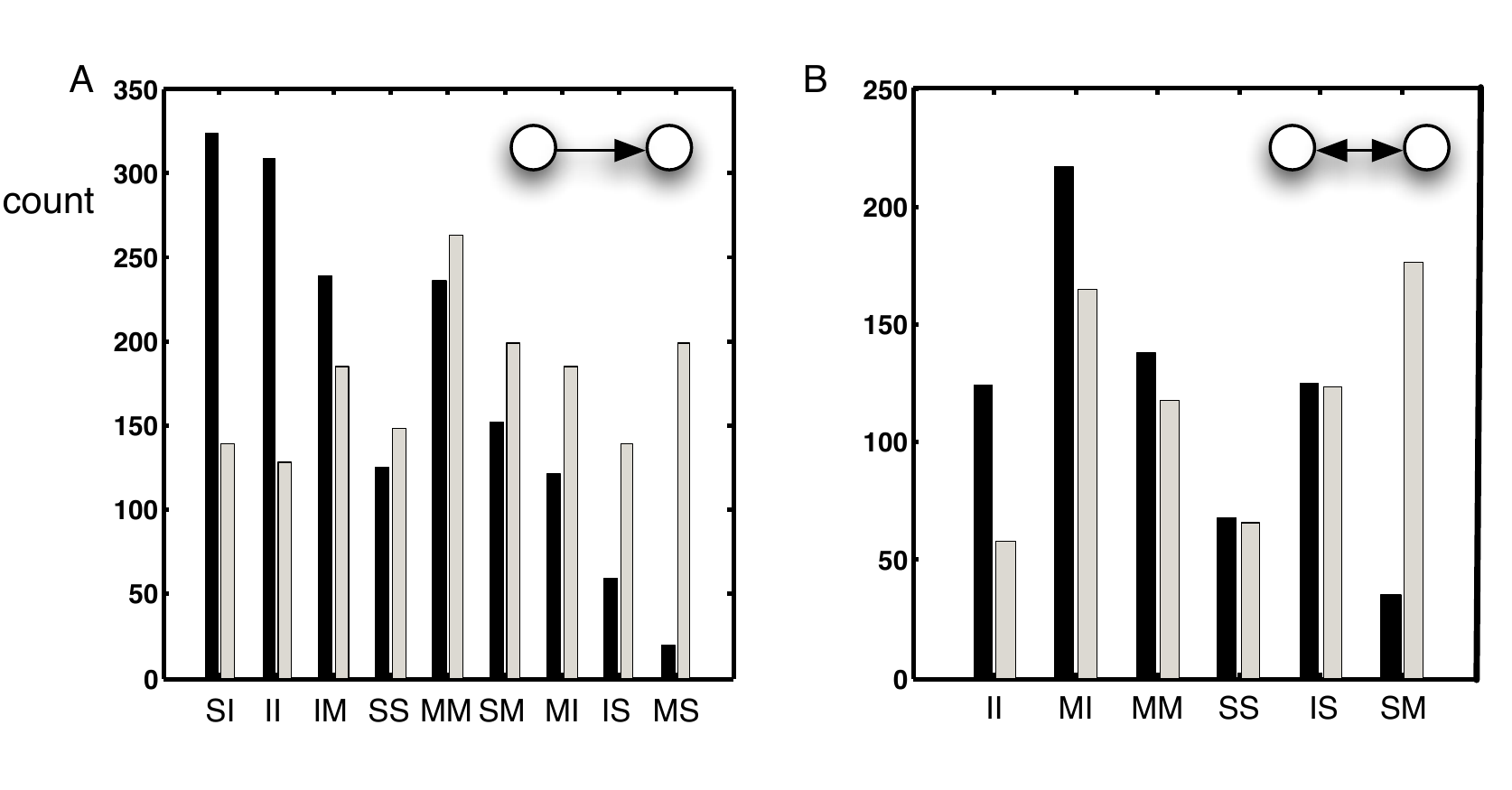} 
   \caption{Histogram of abundances of directed structural motifs with particular coloration in \ce (black) compared to the average abundance in 1,000 color randomizations (grey). S: sensory neuron, I: interneuron, M: motor neuron. A: directed pairs (the direction of information flow is left-to-right: SI means S$\to$I and so forth). B: bi-directional pairs.} 
   \label{fig:hist-2}
\end{figure}

We can study the functional constraints imposed on motifs by node function (color) by analyzing the constraints separately for each of the color realizations of the two motifs. In Fig.~\ref{fig:hist-2} we show the measured counts of each of the color realizations of the directed (Fig.~\ref{fig:hist-2}A) and bi-directional (Fig.~\ref{fig:hist-2}B) motifs. 
%Fig.2
These distributions show that the observed functional constraints make intuitive sense. For example, the ``S$\to$I" (green$\to$red)  as well as ``I$\to$I" (red$\to$red) motifs appear significantly more often then expected by chance, while the motif ``M$\to$S" (blue$\to$green) is significantly suppressed: we do not expect  muscles to relay information to sensory neurons in a functioning worm (even though some of these connections are indeed observed). So, while the uni-directional motif was unremarkable compared to an edge-randomized control, the motifs with ``sensible" colorations such as sensor-$\to$inter-neuron and inter-$\to$inter-neuron are in fact highly significant, while non-sense pairs such as motor-$\to$sensor-neuron are highly unlikely. Such an analysis can reveal motifs in the \ce brain that are used much more frequently than would be expected by chance, which can allow us to dissect the computational building blocks of the network~\cite{Qianetal2011}. 
%Fig. 3
\begin{figure}[!htbp] %  figure placement: here, top, bottom, or page
   \centering
   \includegraphics[width=4.5in]{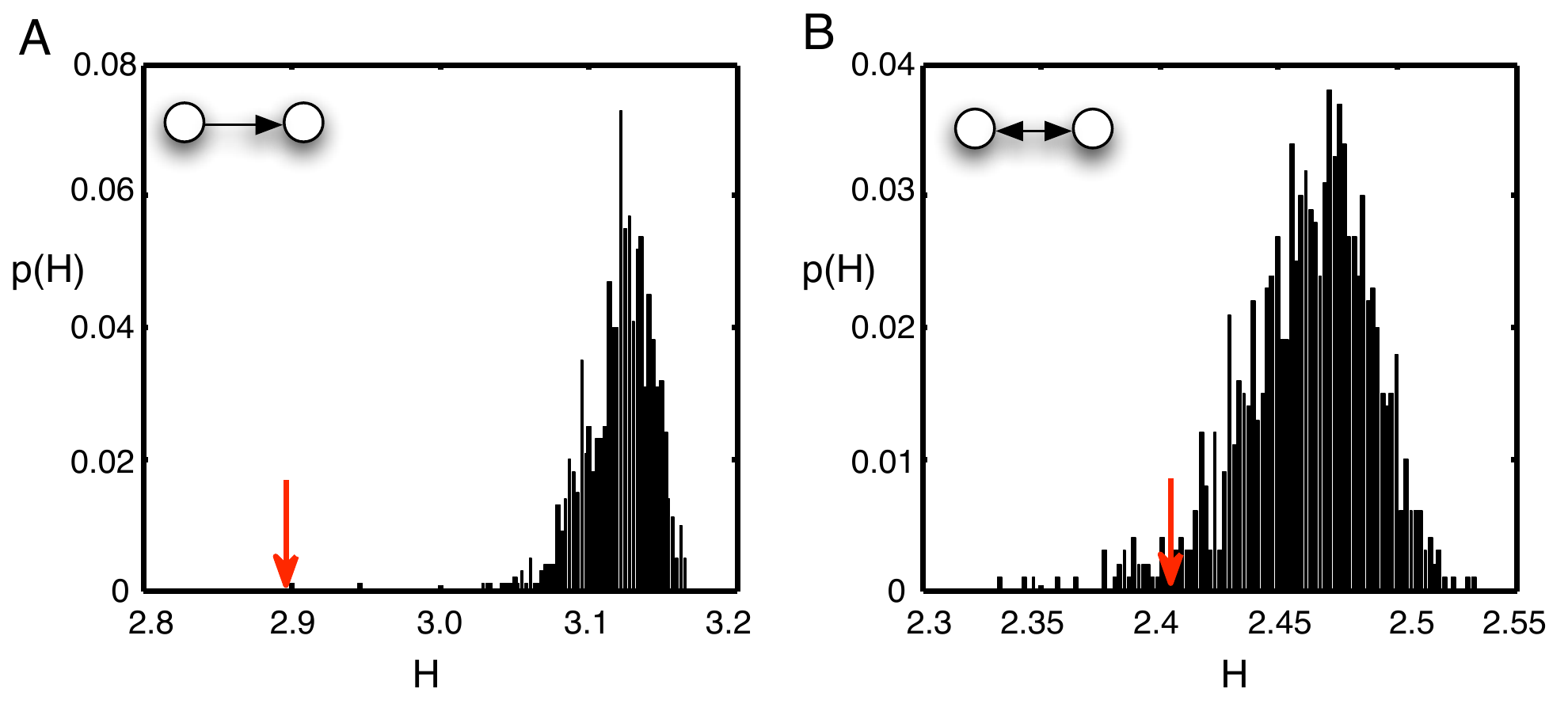} 
   \caption{Distribution of color entropy for the two directed structural motifs with two nodes, obtained from 1,000 color randomizations of the \ce neuronal network. The color entropy of the actual \ce network $H_{\rm col}(X)$ is indicated by the arrow. A: unidirectional two-node motif, B: bi-directional two-node motif.} 
   \label{fig:entropy-2}
\end{figure}

In Fig.~\ref{fig:entropy-2} we compare the color entropy $H_{\rm color}(X)$ of the two structural motifs with two nodes to the distribution of $H^R_{\rm color}(X)$ of 1,000 independent color randomizations of the same network (in color randomizations, the relative count of colors in the network is kept constant). We find that the color entropy of the \ce motifs of two nodes are significantly smaller than their randomized counterparts, a result that is particularly strong for the directed link motif in Fig.~\ref{fig:entropy-2}A. 
Thus, in terms of significant colorations, the uni-directional motif is more remarkable than the bi-directional motif.
From those graphs, we can also estimate the color information content for each motif based on Eq.~(\ref{colinfo}). We find for the information content of the uni-directional motif with two nodes $I_2^{\rm uni}=3.15-2.9=0.35$ bits per symbol, %check exact numbers
while the color-information content of the bi-directional motif is significantly less: $I_2^{\rm bi}=2.47-2.41=0.06$ bits per symbol.
%Fig. 4
\begin{figure}[!htbp] %  figure placement: here, top, bottom, or page
   \centering
   \includegraphics[width=6in]{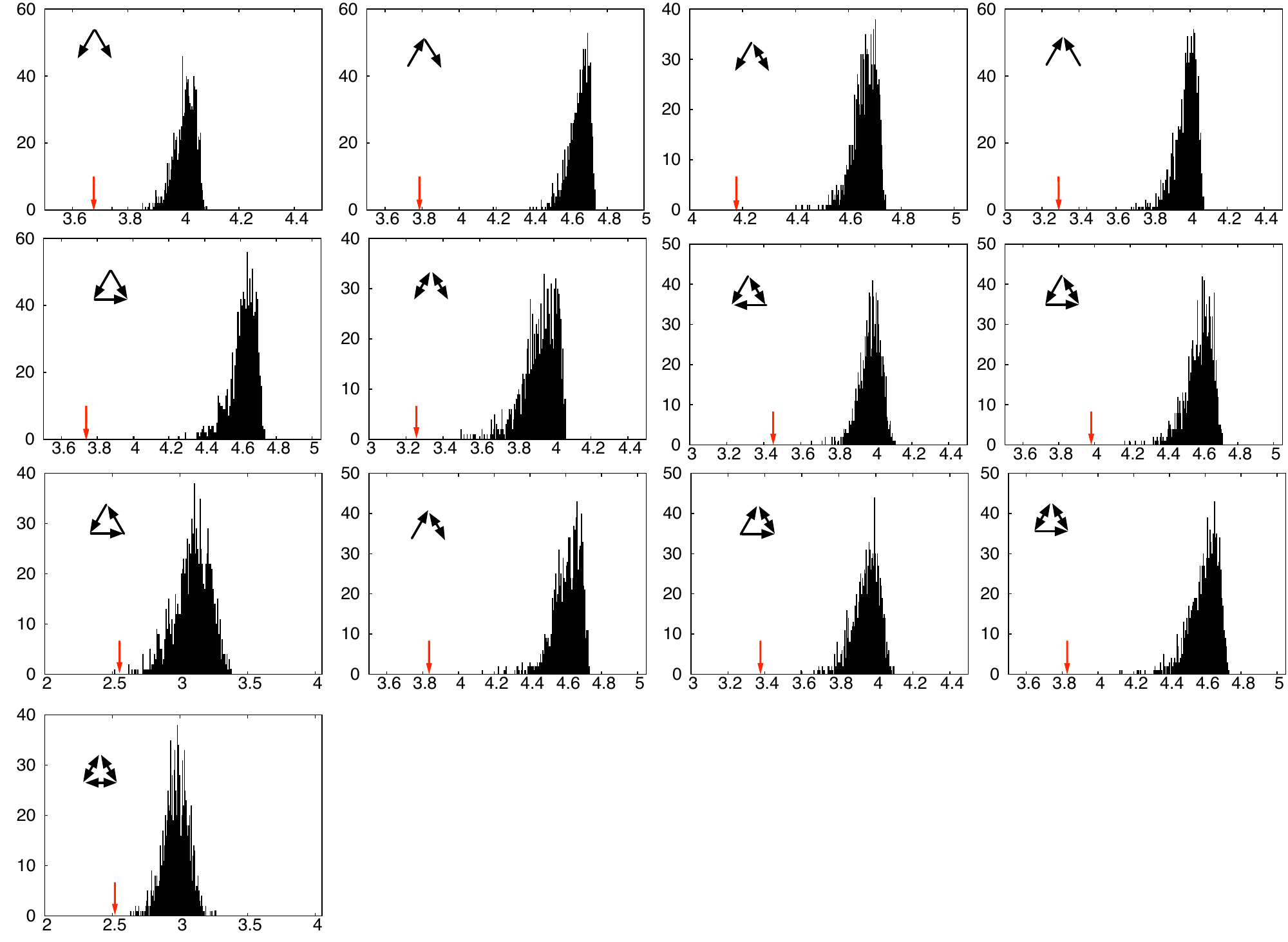} 
   \caption{Distribution of color entropies for the 13 topologically different motifs of directed graphs of three nodes. The motif itself is identified in the upper left corner of each of the 13 histograms, which show the entropy on the $x$-axis and the count of how many times this entropy was observed in 1,000 randomizations of the network on the $y$-axis. The arrow indicates the color entropy of the actual \ce version of this motif, which in all cases is significantly lower than any of the entropies of the randomized networks, but the level of significance varies.}
   \label{fig:entropy-3}
\end{figure}

\subsubsection{Three-node motifs}We can repeat the same analysis for motifs of size 3 (see Fig.~\ref{fig:entropy-3}). There are thirteen different structural motifs, whose color entropy can be measured for \ce and compared to randomized-color controls. Fig.~\ref{fig:entropy-3} shows that {\em all} three-node motifs in \ce have exceptional color combinations that reflect strong selective pressures on which motifs make sense within a functioning worm.

\subsubsection{Entropy and information trends}
\begin{figure}[!htbp] %  figure placement: here, top, bottom, or page
   \centering
   \includegraphics[width=3in]{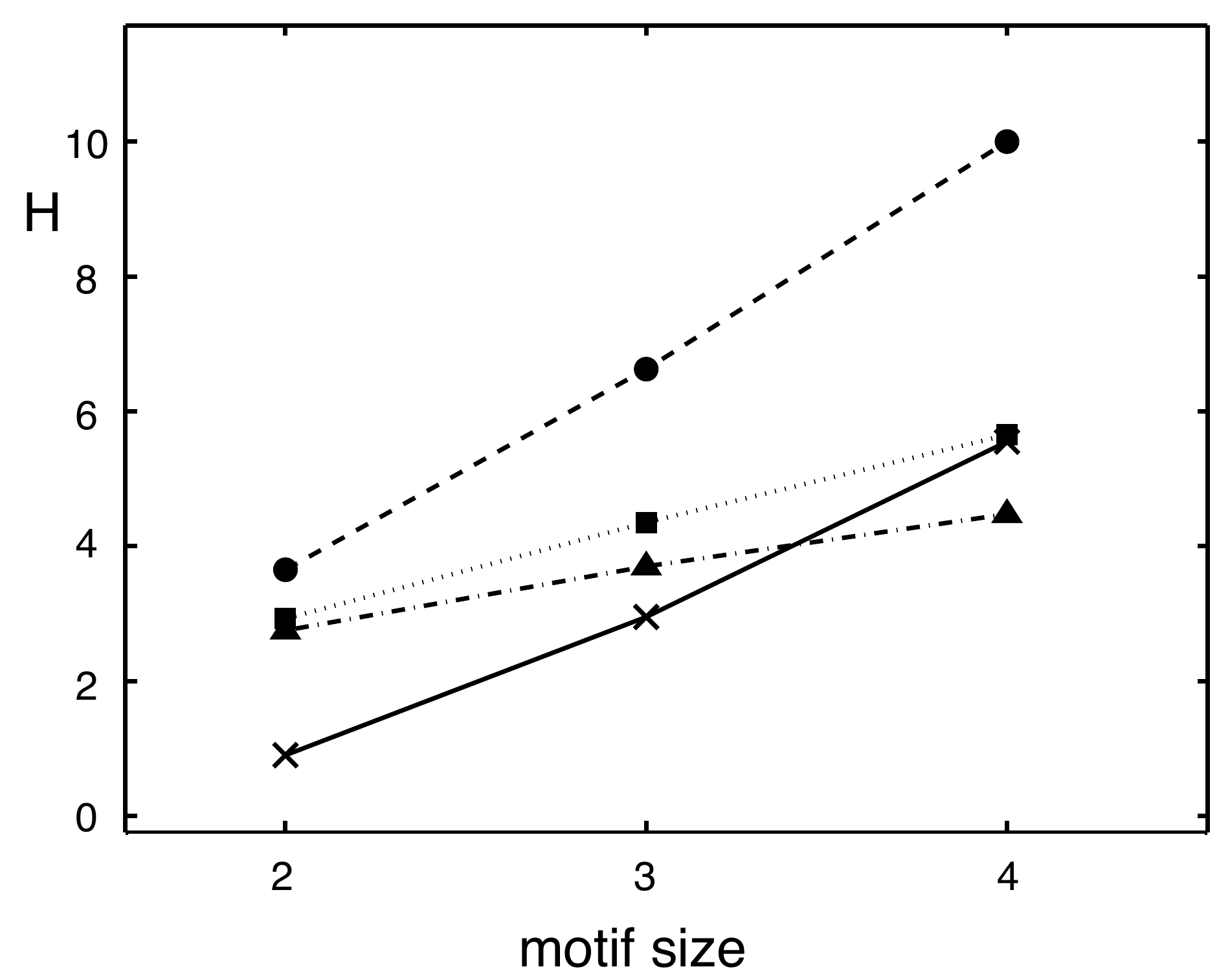} 
   \caption{Motif entropies as a function of motif size. Average color entropy (\ding{115}, dash-dotted line), average randomized color entropy (\ding{110}, dotted), topological entropy (\ding{53}, solid), and total entropy (\ding{108}, dashed). 
   \label{fig:color-ent}}
\end{figure}

Figures \ref{fig:entropy-2} and \ref{fig:entropy-3} indicate that each topological motif has a color entropy that is significantly lower than the average color entropy of that motif in a randomized network. But what is the average color entropy per motif, as a function of motif size? 
The average color entropy is about 2.8 bits for two-node motifs (averaged over the two types studied in Fig.~\ref{fig:entropy-2}), and increases slowly as the number of nodes increases (see Fig.~\ref{fig:color-ent}, dash-dotted line). At the same time, the color entropy for a randomized graph starts at about 2.9 bits per symbol, but increases more quickly, indicating that the amount of functional (that is, color) information per symbol increases from about 0.1 bits per symbol (two-node motifs) to 1.2 bits (four-node motifs).

%Fig. 5

The analysis of information content in structural and colored motifs shows that information is stored in both topology and function, and that the information content depends on the size of the alphabet that is used. At the same time, it is clear that how we assign colors to nodes will also significantly affect information content. For example, other classifications of neuronal functions in the worm exist (such as into ten different morphological classes~\cite{AchacosoYamamoto1992}). However, using more than a handful of colors can quickly make a computational analysis of colored motifs unwieldy because of the explosion in the number of motifs, and we do not expect to see dramatic changes in the information content once a meaningful set of colors is found for a network. 

To study how color information changes as a network evolves, we have to use a different example than the worm brain, as it represents only one snapshot in time. To study motif information evolution, we turn instead to Artificial Life.

\section{Motifs in digital genomes}
In Digital Life~\cite{Ray1992,Adami1998}, populations of self-replicating computer programs are adapting to a user-defined landscape, using a short instruction set of between 20-30 instructions. Since the initial implementation by Tom Ray in the tierra software, most digital life research has been carried out using the Avida platform (see, e.g., the Artificial Life Journal Special Issue~\cite{AdamiWilke2004}, and~\cite{Adami2006}).  Here, we use digital genomes evolved with Avida 2.8.1 (available from SourceForge.net) to create networks of interacting instructions. The 26 instructions in this experiment can be assigned to four different classes of instructions, as shown in Table~\ref{tab:inst}. 
%Tab. 1
\begin{table}[htbp]
   \centering
   %\topcaption{Table captions are better up top} % requires the topcapt package
   \begin{tabular}{@{} cccc @{}} % Column formatting, @{} suppresses leading/trailing space
      \toprule
          Reproductive (Black)   &  Computational (Green) & Flow Control (Blue) & No-Ops (Red) \\
      \midrule
      {\tt h-divide}    & {\tt IO} & {\tt set-flow} & {\tt nop-A} \\
      {\tt h-copy}      & {\tt nand}& {\tt if-less} & {\tt nop-B}\\
    {\tt h-alloc}      & {\tt swap}& {\tt if-label} & {\tt nop-C}\\
                      & {\tt shift-l}& {\tt get-head} & \\
                      & {\tt shift-r}& {\tt mov-head} & \\
                      & {\tt push}&  {\tt jump-head}& \\
                      & {\tt swap-stk}& {\tt mov-head} & \\
                      & {\tt pop}& {\tt if-n-equ} & \\
                      & {\tt add}& & \\
                      & {\tt sub}&  & \\
                      &{\tt inc}& & \\
                      &{\tt dec}& & \\   
       \bottomrule
   \end{tabular}
   \caption{Functional and color assignment of the 26 Avida instructions. The class of ``reproductive" instructions are involved solely in the management of inheritance, while the ``computational" instructions play the role of ``metabolic" instructions, as they are involved in harnessing the energy that avidians need to reproduce. ``Flow control" instructions manage the information flow in the network, while the No-Op instructions are themselves inert, but typically modify the instruction (or instructions) just preceding it.}
   \label{tab:inst}
\end{table}

\begin{figure}[!htbp] %  figure placement: here, top, bottom, or page
   \centering
   \includegraphics[width=4in]{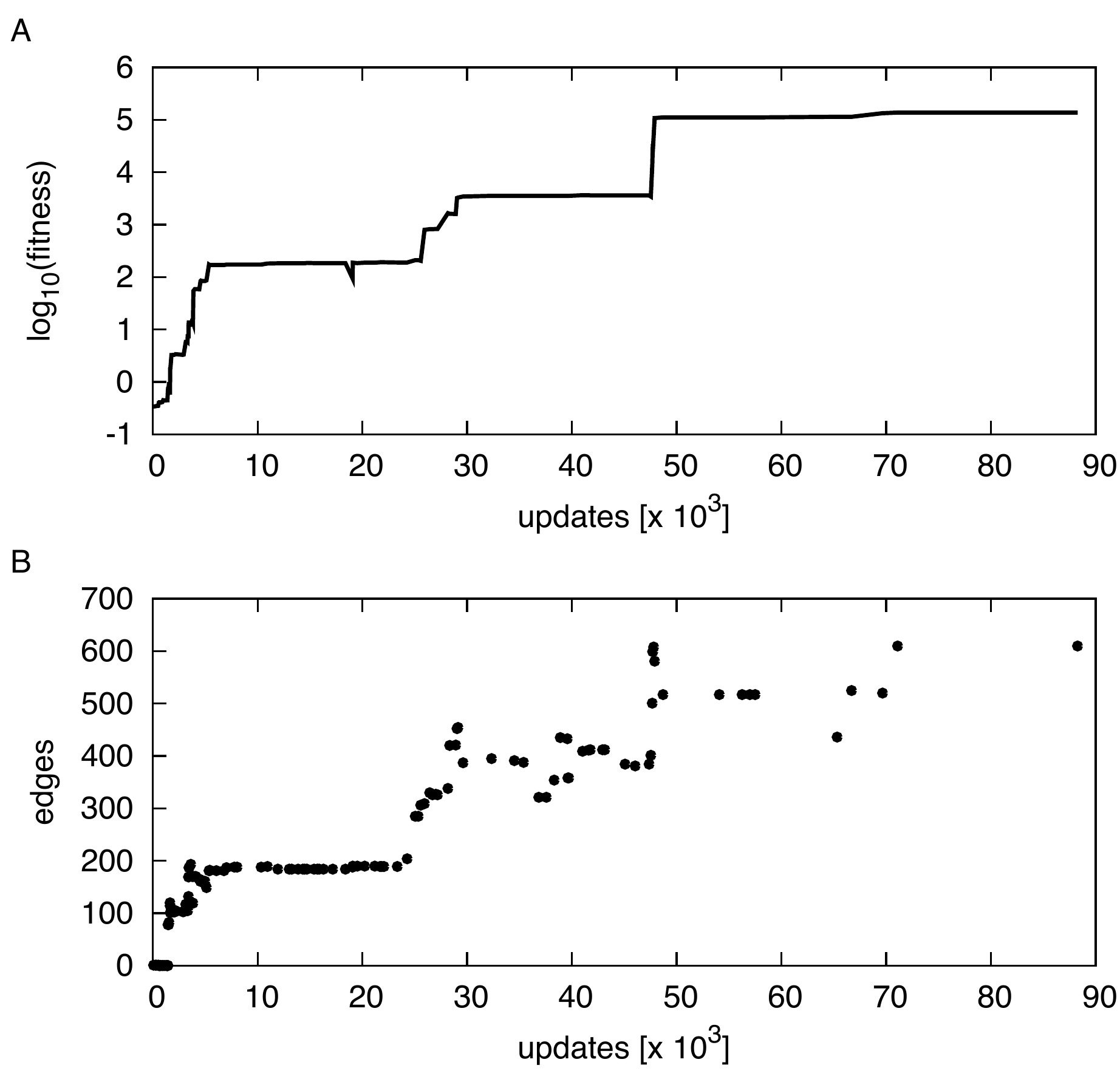} 
   \caption{Fitness, and number of epistatic edges on the line of descent. A: Fitness for 138 genotypes on the LOD. B: Number of edges between interacting instructions with epistasis $|\epsilon_{ij}|>2$, on the LOD.  }
   \label{fig-avida}
\end{figure}

We evolve genomes in the standard ``logic task" landscape, which rewards the performance of all one-input and two-input logical tasks with bonus CPU time depending on the difficulty of the task (there are nine distinct such tasks, see, e.g.,~\cite{Adami2006}). The experiment is started with a population of 3,600 ancestral genomes with a length of 50 instructions that are only capable of self-replication (the self-replicating sequence is padded with the {\tt nop-C} instruction to arrive at the sequence length of 50). The population evolves for 100,000 updates (a measure of time within which each sequence in the population has 30 of its instructions executed), with a mutation rate of 0.0025 per instruction per copy-event (no cross-over), and an instruction-insert and delete probability of 5\% in mass-action mode (well-mixed chemostat). Because of the insert/delete probability, the sequence length is not constant, but instead increases slightly during evolution to 56 instructions. Fig.~\ref{fig-avida}A shows the evolution of fitness as a function of updates, on the {\em line of descent} (LOD) of the population. The line of descent is created by picking a representative of the most fit genotype of the population at the end of the experiment, and tracing its lineage backwards in time via its direct ancestors, ending at the seed genotype. Because these populations evolve in a single niche, the LODs of all genotypes present in the population at the end of the experiment quickly coalesce, so that a single LOD characterizes the evolutionary dynamics of the experiment~\cite{Lenskietal2003}. Analysis of evolutionary experiments in terms of the LOD (rather than population averages) has the advantage of recapitulating the salient events in evolutionary history, while disregarding any changes that did not leave a trace in the final product. In that manner, the LOD allows for a reconstruction of the path that evolution took to arrive at the adapted sequence.
%Fig. 6
\subsection{Epistasis}
We determine that two instructions within an avidian sequence interact if the fitness effects of knocking out these instructions depend on each other, that is, if the fitness contribution of one instruction is contingent on the identity of the other. Two instructions that are linked in such a way are called {\em epistatic}~\cite{Wolfetal2000} (see also the review~\cite{Phillips2008}). Epistasis is an important concept within evolutionary genetics, and is usually defined to quantify the interaction between genes~\cite{Ostmanetal2011} rather than between the set of monomers that code for the gene, but can easily be used the way we show here~\cite{Lenskietal1999}. It has been shown earlier that complex genomes show more positive epistasis between deleterious mutations than simple ones~\cite{Lenskietal1999} (a finding we corroborate here), and that epistasis between avidian instructions is crucial to understand the evolution of complex features~\cite{Lenskietal2003}.

Instruction knockouts are performed by replacing each instruction by an inert instruction {\tt nop-X}, in order to prevent fitness effects that are due to a change in sequence length only rather than the identity of the instruction. For each sequence on the LOD, we can calculate the epistasis $\epsilon_{ij}$ for any pair of mutations at instruction sites $i$ and $j$ as follows. Let the unmutated (that is wild-type) fitness of the sequence be $w_0$. Here, fitness is measured as the rate at which an avidian produces offspring per generation, and is equivalent to the growth rate of more conventional organisms. The fitness effect of mutating instruction $i$ then is $w_i/w_0$. On the LOD, we find many substitutions that are neutral or beneficial, however, most knockouts of arbitrary instructions are either neutral or deleterious. After creating the mutant with fitness $w_i$, mutate another instruction $j$ to obtain the double-mutant with fitness $w_{ij}$. At the same time, revert mutation $i$ on the double mutant to obtain a genome with only the single mutation $j$, with fitness $w_j$. This is sufficient to compare the two single-mutant effects $w_i/w_0$ and $w_j/w_0$ with the effect of the double mutant $w_{ij}/w_0$. The quantity
\be
\epsilon_{ij}=\log_e\left(\frac{w_{ij} w_0}{w_i w_j}\right)
\ee
then measures the epistasis between the two instructions $i$ and $j$ (see, e.g.,~\cite{Ostmanetal2011} and references therein). Positive epistasis between mutations implies that the fitness of the double mutant is higher than we would have expected from the effect of each single mutation, while negative epistasis signifies that the double mutation has made things worse than either of the single mutations would have led you to believe. A typical example of genetic (epistatic) interaction is a pair of redundant instructions, where each of the mutations by themselves does not affect organism fitness, while the mutation of both instructions creates a fitness deficit. In this case, the epistasis is clearly negative. This effect is called ``synthetic lethality" (if the double mutant is non-viable) in the genetic literature. The opposite case can also occur, when the knockout of one gene compensates for the loss of function due to the knockout of another, but the effect is less common. In general, more interactions in avidians are of the positive sort~\cite{Lenskietal1999}, simply because a second mutation that affects the same functional block as the first has virtually no effect anymore. As a consequence, groups of epistatically connected instructions often outline functional blocks or modules.

\begin{figure}[!htbp] %  figure placement: here, top, bottom, or page
   \centering
   \includegraphics[width=5in]{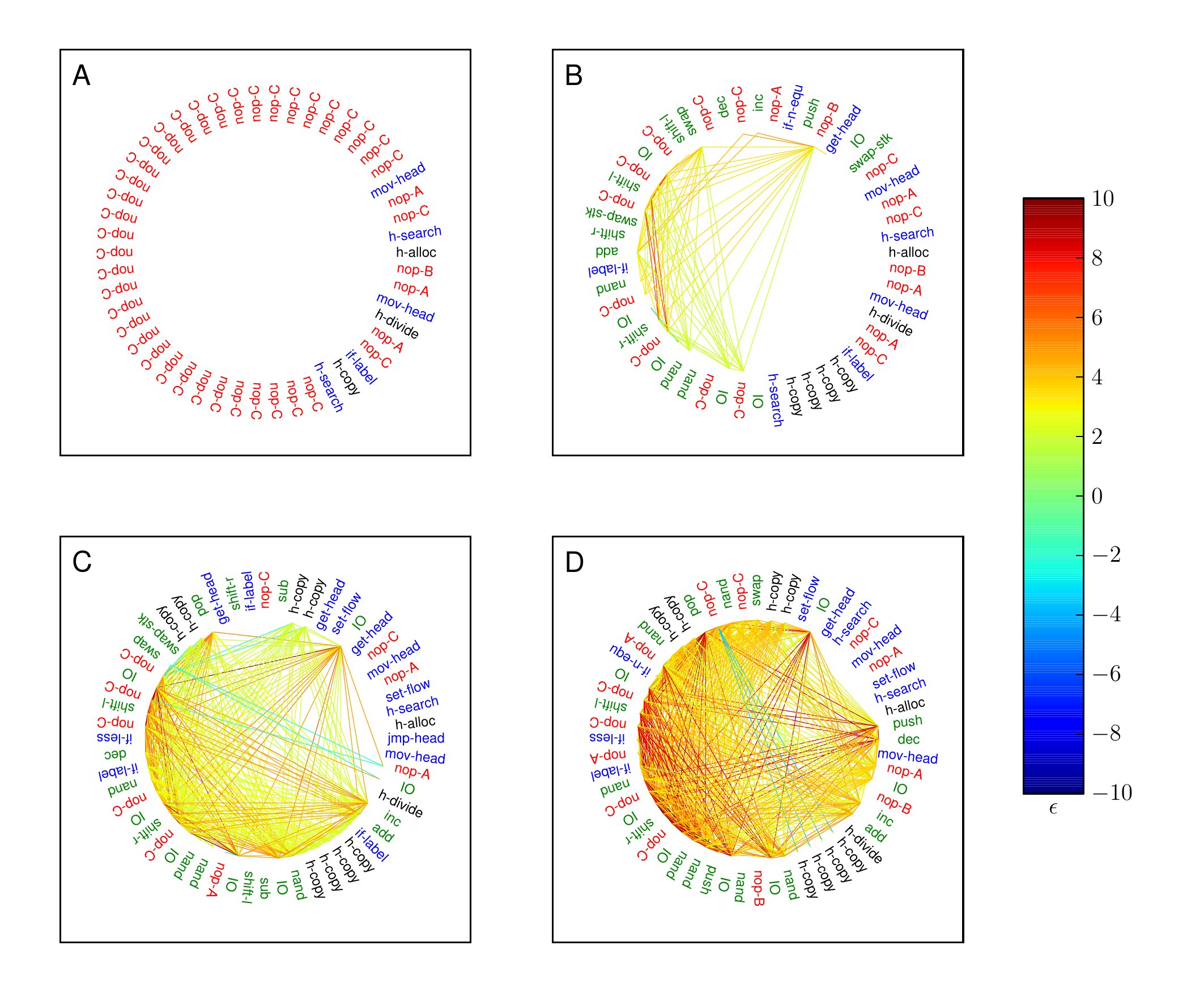} 
   \caption{Epistatic interactions between instructions for four genomes on the LOD. Instructions are colored according to the scheme detailed in Table 1, while epistatic edges are colored according to their strength and direction, in a graded manner between blue ($\epsilon=-10$) over green (vanishing epistasis, not shown in these plots because of the threshold) to red ($\epsilon=10$, see color bar). A: Ancestral genome (50 instructions). B: A genome early on the LOD (at update 3,742). C: Genotype on the LOD at update 29,035. D: Epistatic network for the last genotype on the LOD, at update 88,297, with 610 edges.}
   \label{fig-ko}
\end{figure}

In Fig.~\ref{fig-ko}, we see avidian sequences at different time points in their evolution, with instructions colored according to the functional tags defined in Table~\ref{tab:inst}, and edges indicating epistatic interactions for pairs of instructions $i$ and $j$ if $|\epsilon_{ij}|>2$. With this cutoff, there are no interactions between instructions in the ancestral genotype (Fig.~\ref{fig-ko}A), but they start to emerge around update 1,450. Note that even though the fitness rises exponentially, epistasis is defined in terms of the relative effect of a knockout on fitness, and we should not expect a priori that the number of edges should increase as fitness increases. The cutoff $|\epsilon_{ij}|>2$ is quite stringent: it implies that the double mutants fitness effect must be more than ${\rm e}^2$ times the product of the fitness effects of the single mutations (for positive epistasis), or less than $\approx13.5\%$ of the product (for negative epistasis).
%Fig. 7

\subsection{Motif entropy and information}
\begin{figure}[!htbp] %  figure placement: here, top, bottom, or page
   \centering
   \includegraphics[width=4in]{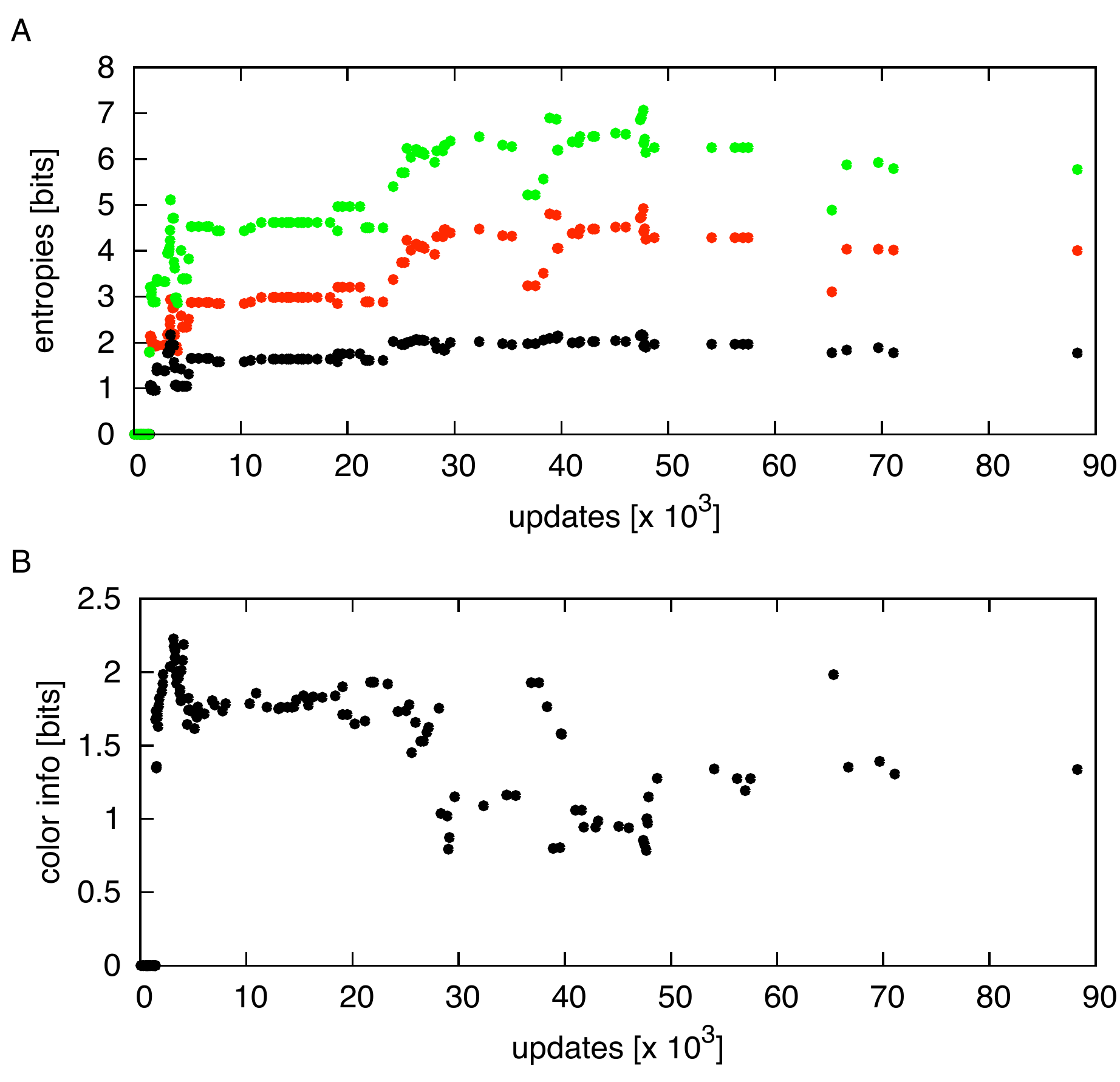} 
   \caption{Motif entropies and colored motif information content. A: Motif entropies for the 138 genotypes on the LOD. Black dots: topological entropy of motifs of size 4 (maximal entropy for 6 motifs is 2.585 bits). Red dots: average color entropy of motifs of size 4. Green dots: Total motif entropy [Eq.~(\ref{tot_ent})], given by the sum of topological and color entropy, according to the grouping axiom. B: Information content (per motif) in colored motifs of size 4, according to Eq.~(\ref{colinfo}).}
   \label{fig-entropy}
\end{figure}

For the epistatic colored networks of the type shown in Fig.~\ref{fig-ko}, we can calculate colored motif entropies, structural motif entropies, and colored motif information content as described above. We focus here on motifs of size four, which are undirected. There are six structural motifs of size four (shown in Fig.~\ref{colmotifs}C), which come in a total of 566 different colorations. Thus, the maximal total motif entropy is $H_{\rm tot}^{\rm max}\approx 9.14$ bits. We count the frequency of colored motifs in the network as described for the {\it C. elegans} network, and also calculate the mean frequency with which we observe that motif in a color-randomized network. In Fig.~\ref{fig-entropy}A, we show the topological entropy calculated using Eq.~(\ref{topent}), the color entropy [Eq.~(\ref{colentav})], as well as the total entropy
[from Eq.~(\ref{tot_ent})], along the LOD of the experiment. Generally, entropies are increasing as the network evolves, not because the network increases in size but because the number of edges is increasing (as seen in Fig.~\ref{fig-avida}B), which leads to a greater diversity of colored motifs. However, the genetic changes that give rise to the fitness jump around 50,000 updates (see Fig.~\ref{fig-avida}A) appear to change the genetic architecture in such a manner that color diversity {\em decreases} somewhat. This decrease is most apparent in the color entropy, less so in the structural motif entropy. 

In order to calculate the information stored in the color assignment of instructions, we need to calculate the average color entropy for color randomized networks. We do this as described above for the {\it C. elegans} motifs, to obtain $\la H_{\rm color}^R\ra$. This entropy also increases with evolutionary time, and as a consequence the difference between the two is mostly constant, but also shows a decrease at least for some periods of time on the LOD (see Fig.~\ref{fig-entropy}). It is not immediately clear what kind of changes in the genetic architecture of the sequences is responsible for the drop or increase in motif information content. However, because the network is comparatively small, small changes in the genome can potentially give rise to large changes in the colored motif distribution. Note that the color entropy for 1000 color-randomized networks has an error in the mean that is much smaller than the changes seen on the LOD (between 0.02 for the earliest networks to 0.005 for the fittest ones), indicating that the fluctuations are not due to sampling error. We conclude that the color assignment (shown in Table 1) that we chose for the instructions shows that some information is stored within the colored motifs in the epistatic network, but that this information does not necessarily increase with an increase in fitness. In particular, it is possible that a different choice of color assignments captures {\em more} motif information, and correlates differently with fitness. Thus, while the genomic information content~\cite{AdamiCerf2000,Adamietal2000,Adami2004} correlates very well with fitness (see also~\cite{Huangetal2004}), the colored motif information content appears to be better suited to track changes in the genome architecture and organization. 

%Fig. 8
\section{Discussion}
We have shown how an information-theoretic analysis of networks in which nodes are assigned a color based on their functionality allows us to determine the information content of the network motifs in a manner that significantly expands the purely topological treatment. The method is general and can be applied to any network where both structural information (connectivity) and functional annotation of the nodes is available.
When considering the neuronal network of \ce as an example, we note that (depending on what size motifs we consider) more information is stored in the coloration of the motifs than in the structure. Indeed, an analysis of the \ce brain in terms of structural motifs has generated only limited insight~\cite{Reigletal2004,Songetal2005}, while adding the color degree of freedom creates a wealth of information about what computations are performed by the worm's brain~\cite{Qianetal2011}. An analysis of the information stored in motif colorations shows that the information per symbol increases with the size of the motifs considered, but while it is clear that this information is a consequence of selection (because it is precisely the difference between a random color entropy and that selected by evolution) it is not clear how this information changes as the organism adapts. To address this question, we have analyzed the information content of colored motifs in networks created by the interaction between instructions of avidian genomes. By choosing a particular functional coloring of instructions (here 4 colors tagging instructions that have either a biological, a computational, a flow-control, or a modifying function), we discover that while information is stored in the colorations, this information neither has to increase nor decrease with adaptation. While the number of epistatic edges increase as the organism adapts to its environment, the colored motif distribution (while clearly constrained by the functionality of the sequence) can become more narrow or more broad, depending on the genetic architecture of the sequence that gives rise to them. We do see clear indications that the distribution changes at stages in which new functionality is evolved, which points to a relation between genomic architecture and colored motif distribution, but monitoring the information content alone is not sufficient to dissect what these changes are. 

Of course, no general conclusions about the evolution of colored motif information in complex networks can be drawn from this single example, not because a single experiment would not be reflective of the average evolutionary trajectory (we believe it is in the present case) but rather because the assignment of colors to functions of the instructions reflects the investigator's intuition, but is not necessarily the assignment that maximizes the information content. Thus, while it is clear from the example we studied here that information content of colored motifs cannot be a universal measure of network complexity independently of what the color assignment is (or how edges are defined), it is nevertheless a promising tool for dissecting the functional complexity of a network. It is interesting to ask whether a search over possible colorations (using a limited number of colors, of course) looking for that coloration that maximizes the information content of motifs could generate insight into the functionality of instructions (and their dependence) that is not obvious from the outset. 

\section{Acknowledgements}
We would like to thank Charles Ofria and Bj{\o}rn {\O}stman for discussions. This work was supported in part by the National Science Foundation's BEACON Center for the Study of Evolution in Action, under contract No. DBI-0939454 and by the Agriculture and Food Research Initiative Competitive Grant no. 2010-65205-20361 from the USDA National Institute of Food and Agriculture. We wish to acknowledge the support of the Michigan State University High Performance Computing Center and the Institute for Cyber Enabled Research.

\bibliographystyle{alife_mit}
\bibliography{Adami}

\begin{thebibliography}{10}

\bibitem{AchacosoYamamoto1992}
Achacoso, T. \& Yamamoto, W. (1992).
\newblock \emph{AY's Neuroanatomy of {\it C. elegans} for computation}.
\newblock Boca Raton: CRC Press.

\bibitem{Adami1998}
Adami, C. (1998).
\newblock \emph{Introduction to Artificial Life}.
\newblock New York: Springer Verlag.

\bibitem{Adami2002b}
Adami, C. (2002).
\newblock What is complexity?
\newblock \emph{BioEssays}, \emph{24}, 1085--94.

\bibitem{Adami2004}
Adami, C. (2004).
\newblock Information theory in molecular biology.
\newblock \emph{Physics of Life Reviews}, \emph{1}, 3--22.

\bibitem{Adami2006}
Adami, C. (2006).
\newblock Digital genetics: unravelling the genetic basis of evolution.
\newblock \emph{Nature Reviews Genetics}, \emph{7}, 109--118.

\bibitem{AdamiCerf2000}
Adami, C. \& Cerf, N.~J. (2000).
\newblock Physical complexity of symbolic sequences.
\newblock \emph{Physica D}, \emph{137}, 62--69.

\bibitem{Adamietal2000}
Adami, C., Ofria, C., \& Collier, T. (1999).
\newblock Evolution of biological complexity.
\newblock \emph{Proc. Natl. Acad. Sci. USA}, \emph{97}, 4463--4468.

\bibitem{AdamiWilke2004}
Adami, C. \& Wilke, C. (2004).
\newblock Experiments in digital evolution ({E}ditors' introduction to the
  special issue).
\newblock \emph{Artificial Life}, \emph{10}, 117--122.

\bibitem{Ahnertetal2010}
Ahnert, S.~E., Johnston, I.~G., Fink, T. M.~A., Doye, J. P.~K., \& Louis, A.~A.
  (2010).
\newblock Self-assembly, modularity, and physical complexity.
\newblock \emph{Physical Review E}, \emph{82}, 026117.

\bibitem{AlbertBarabasi2002}
Albert, R. \& Barabasi, A.-L. (2002).
\newblock Statistical mechanics of complex networks.
\newblock \emph{Reviews of Modern Physics}, \emph{74}, 47--97.

\bibitem{Ash1965}
Ash, R.~B. (1965).
\newblock \emph{Information Theory}.
\newblock New York, N.Y.: Dover Publications, Inc.

\bibitem{Bennett1995}
Bennett, C. (1995).
\newblock Universal computation and physical dynamics.
\newblock \emph{Physica D}, \emph{86}, 268--273.

\bibitem{Bianconi2008}
Bianconi, G. (2008).
\newblock The entropy of randomized network ensembles.
\newblock \emph{Europhysics Letters}, \emph{81}, 28005.

\bibitem{Carothersetal2004}
Carothers, J.~M., Oestreich, S.~C., Davis, J.~H., \& Szostak, J.~W. (2004).
\newblock Informational complexity and functional activity of {RNA} structures.
\newblock \emph{J. American Chem. Society}, \emph{126}, 5130--5137.

\bibitem{Claussen2007}
Claussen, J.~C. (2007).
\newblock Offdiagonal complexity: A computationally quick complexity measure
  for graphs and networks.
\newblock \emph{Physica A}, \emph{375}, 365--373.

\bibitem{CoverThomas1991}
Cover, T.~M. \& Thomas, J.~A. (1991).
\newblock \emph{Elements of Information Theory}.
\newblock New York, NY: John Wiley.

\bibitem{Dornetal2011}
Dorn, E.~D., Nealson, K.~H., \& Adami, C. (2011).
\newblock Monomer abundance patterns as a universal biosignature: Examples from
  terrestrial and artificial life.
\newblock \emph{Journal of Molecular Evolution}, \emph{72}, 283--295.

\bibitem{EbelingJimenezMontano1980}
Ebeling, W. \& Jimenez-Montano, M. (1980).
\newblock On grammars, complexity, and information measures of biological
  macromolecules.
\newblock \emph{Mathematical Biosciences}, \emph{52}, 53--71.

\bibitem{GellMannLloyd1996}
Gell-Mann, M. \& Lloyd, S. (1996).
\newblock Information measures, effective complexity, and total information.
\newblock \emph{Complexity}, \emph{2}, 44--52.

\bibitem{HallRussell1991}
Hall, D. \& Russell, R. (1991).
\newblock The posterior nervous system of the nematode caenorhabditis elegans:
  Serial reconstruction of identified neurons and complete pattern of synaptic
  interactions.
\newblock \emph{J. Neuroscience}, \emph{11}, 1--22.

\bibitem{Hazenetal2007}
Hazen, R.~M., Griffin, P.~L., Carothers, J.~M., \& Szostak, J.~W. (2007).
\newblock Functional information and the emergence of biocomplexity.
\newblock \emph{Proc Natl Acad Sci U S A}, \emph{104 Suppl 1}, 8574--81.

\bibitem{HintzeAdami2008}
Hintze, A. \& Adami, C. (2008).
\newblock Evolution of complex modular biological networks.
\newblock \emph{PLoS Computational Biology}, \emph{4}, e23.

\bibitem{HintzeAdami2010}
Hintze, A. \& Adami, C. (2010).
\newblock Modularity and anti-modularity in networks with arbitrary degree
  distribution.
\newblock \emph{Biol Direct}, \emph{5}, 32.

\bibitem{Huangetal2004}
Huang, W., Ofria, C., \& Torng, E. (2004).
\newblock Measuring biological complexity in digital organisms.
\newblock In J.~Pollack, M.~A. Bedau, P.~Husbands, T.~Ikegami, \& R.~Watson
  (Eds.) \emph{Proceedings of Artificial Life IX}. Cambridge, MA: MIT Press,
  (pp. 315--321).

\bibitem{KimWilhelm2008}
Kim, J. \& Wilhelm, T. (2008).
\newblock What is a complex graph?
\newblock \emph{Physica A}, \emph{387}, 2637--2652.

\bibitem{Kolmogorov1965}
Kolmogorov, A. (1965).
\newblock Three approaches to the quantitative definition of information.
\newblock \emph{Problems of Information Transmission}, \emph{1}, 4.

\bibitem{LempelZiv1976}
Lempel, A. \& Ziv, J. (1976).
\newblock On the complexity of finite sequences.
\newblock \emph{I{EEE} Transactions On Information Theory}, \emph{22}, 75--81.

\bibitem{Lenskietal1999}
Lenski, R.~E., Ofria, C., Collier, T.~C., \& Adami, C. (1999).
\newblock Genome complexity, robustness and genetic interactions in digital
  organisms.
\newblock \emph{Nature}, \emph{400}, 661--664.

\bibitem{Lenskietal2003}
Lenski, R.~E., Ofria, C., Pennock, R.~T., \& Adami, C. (2003).
\newblock The evolutionary origin of complex features.
\newblock \emph{Nature}, \emph{423}, 139--44.

\bibitem{LiVitanyi1997}
Li, M. \& Vitanyi, P. (1997).
\newblock \emph{An introduction to Kolmogorov complexity and its applications}.
\newblock New York, NY: Springer Verlag.

\bibitem{Lloyd2001}
Lloyd, S. (2001).
\newblock Measures of complexity: A nonexhaustive list.
\newblock \emph{{IEEE} Control Systems Magazine}, \emph{21}, 7--8.

\bibitem{Lofgren1977}
L{\"o}fgren, L. (1977).
\newblock Complexity of description of systems: A foundational study.
\newblock \emph{Int. J. Gen. Sys.}, \emph{3}, 197--214.

\bibitem{McShea2000}
McShea, D.~W. (2000).
\newblock Functional complexity in organisms: Parts as proxies.
\newblock \emph{Biology and Philosophy}, \emph{15}, 641--668.

\bibitem{MeyerOrtmanns2004}
Meyer-Ortmanns, H. (2004).
\newblock Functional complexity measure for networks.
\newblock \emph{Physica A}, \emph{337}, 679--690.

\bibitem{Miloetal2004}
Milo, R., Itzkovitz, S., Kashtan, N., Levitt, R., Shen-Orr, S., Ayzenshtat, I.,
  Sheffer, M., \& Alon, U. (2004).
\newblock Superfamilies of evolved and designed networks.
\newblock \emph{Science}, \emph{303}, 1538--42.

\bibitem{Miloetal2002}
Milo, R., Shen-Orr, S., Itzkovitz, S., Kashtan, N., Chklovskii, D., \& Alon, U.
  (2002).
\newblock Network motifs: simple building blocks of complex networks.
\newblock \emph{Science}, \emph{298}, 824--7.

\bibitem{Newmanetal2006}
Newman, M., Barabasi, A.-L., \& Watts, D. (2006).
\newblock \emph{The Structure and Dynamics of Networks}.
\newblock Princeton, N.J.: Princeton University Press.

\bibitem{Newman2003}
Newman, M. E.~J. (2003).
\newblock The structure and function of complex networks.
\newblock \emph{SIAM Review}, \emph{45}, 167--256.

\bibitem{Ofriaetal2008}
Ofria, C., Huang, W., \& Torng, E. (2008).
\newblock On the gradual evolution of complexity and the sudden emergence of
  complex features.
\newblock \emph{Artif Life}, \emph{14}, 255--63.

\bibitem{Ostmanetal2011}
{\O}stman, B., Hintze, A., \& Adami, C. (2011).
\newblock Impact of epistasis and pleiotropy on evolutionary adaptation.
\newblock Preprint arXiv:0909.3506 on arxiv.org.

\bibitem{Papentin1980}
Papentin, F. (1980).
\newblock On order and complexity {I}: {General} considerations.
\newblock \emph{J. theor. Biol.}, \emph{87}, 421--456.

\bibitem{Papentin1982}
Papentin, F. (1982).
\newblock {On order and complexity II: Application to chemical} and biochemical
  structures.
\newblock \emph{J. theor. Biol.}, \emph{95}, 225--245.

\bibitem{Phillips2008}
Phillips, P.~C. (2008).
\newblock Epistasis - the essential role of gene interactions in the structure
  and evolution of genetic systems.
\newblock \emph{Nature Reviews Genetics}, \emph{9}, 855--867.

\bibitem{Piraveenanetal2010}
Piraveenan, M., Prokopenko, M., \& Zomaya, A. (2010).
\newblock Assortative mixing in directed biological networks.
\newblock \emph{IEEE/ACM Trans Comput Biol Bioinform}.

\bibitem{Qianetal2011}
Qian, J., Hintze, A., \& Adami, C. (2011).
\newblock Colored motifs reveal computational building blocks in the \it {C.
  elegans} \rm brain.
\newblock \emph{PLoS ONE}, \emph{6}, e17013.

\bibitem{Ray1992}
Ray, T.~S. (1992).
\newblock An approach to the synthesis of life.
\newblock In C.~G. Langton, J.~D. Farmer, \& S.~Rasmussen (Eds.)
  \emph{Artificial Life II}. Redwood City: Addison-Wesley, (pp. 371--408).

\bibitem{Reigletal2004}
Reigl, M., Alon, U., \& Chklovskii, D.~B. (2004).
\newblock Search for computational modules in the {C}. elegans brain.
\newblock \emph{BMC Biol}, \emph{2}, 25.

\bibitem{Riceetal2005}
Rice, J.~J., Kershenbaum, A., \& Stolovitzky, G. (2005).
\newblock Lasting impressions: Motifs in protein-protein maps may provide
  footprints of evolutionary events.
\newblock \emph{Proc Natl Acad Sci U S A}, \emph{102}, 3173--4.

\bibitem{Shannon1948}
Shannon, C. (1948).
\newblock A mathematical theory of communication.
\newblock \emph{Bell System Technical Journal}, \emph{27}, 379--423,623--656.

\bibitem{Shannon1951}
Shannon, C.~E. (1951).
\newblock Prediction and entropy of printed {E}nglish.
\newblock \emph{Bell System Technical Journal}, \emph{30}, 50--64.

\bibitem{ShenOrretal2002}
Shen-Orr, S.~S., Milo, R., Mangan, S., \& Alon, U. (2002).
\newblock Network motifs in the transcriptional regulation network of
  {Escherichia coli}.
\newblock \emph{Nat Genet}, \emph{31}, 64--68.

\bibitem{SoleValverde2004}
Sol\'{e}, R. \& Valverde, S. (2004).
\newblock Information theory of complex networks: On evolution and
  architectural constraints.
\newblock \emph{Lect. Notes Phys.}, \emph{650}, 189--207.

\bibitem{SoleValverde2006}
Sol{\'e}, R.~V. \& Valverde, S. (2006).
\newblock Are network motifs the spandrels of cellular complexity?
\newblock \emph{Trends Ecol Evol}, \emph{21}, 419--22.

\bibitem{SoloveichikWinfree2006}
Soloveichik, D. \& Winfree, E. (2006).
\newblock Complexity of self-assembled shapes.
\newblock \emph{{SIAM} Journal On Computing}, \emph{36}, 1544--1569.

\bibitem{Songetal2005}
Song, S., Sj{\"o}str{\"o}m, P.~J., Reigl, M., Nelson, S., \& Chklovskii, D.~B.
  (2005).
\newblock Highly nonrandom features of synaptic connectivity in local cortical
  circuits.
\newblock \emph{PLoS Biol}, \emph{3}, e68.

\bibitem{SpornsKoetter2004}
Sporns, O. \& K{\"o}tter, R. (2004).
\newblock Motifs in brain networks.
\newblock \emph{PLoS Biol}, \emph{2}, e369.

\bibitem{Szostak2003}
Szostak, J.~W. (2003).
\newblock Functional information: Molecular messages.
\newblock \emph{Nature}, \emph{423}, 689.

\bibitem{Thomasetal2000}
Thomas, R.~D., Shearman, R.~M., \& Stewart, G.~W. (2000).
\newblock Evolutionary exploitation of design options by the first animals with
  hard skeletons.
\newblock \emph{Science}, \emph{288}, 1239--1242.

\bibitem{ThomasReif1993}
Thomas, R. D.~K. \& Reif, W.-E. (1993).
\newblock The skeleton space: A finite set of organic designs.
\newblock \emph{Evolution}, \emph{47}, 341--360.

\bibitem{Varshneyetal2009}
Varshney, L.~R., Chen, B.~L., Paniagua, E., Halland, D.~H., \& Chklovskii,
  D.~B. (2011).
\newblock Structural properties of the \it {Caenorhabditis elegans} \rm
  neuronal network.
\newblock \emph{PLoS Computational Biology}, \emph{7}, e1001066.

\bibitem{Whiteetal1986}
White, J., Southgate, E., Thomson, J., \& Brenner, S. (1986).
\newblock The structure of the nervous system of the nematode {C}aenorhabditis
  elegans.
\newblock \emph{Philos Trans R Soc Lond Biol Sci}, \emph{314}, 1--340.

\bibitem{Wilhelm2003}
Wilhelm, T. (2003).
\newblock An elementary dynamic model for non-binary food webs.
\newblock \emph{Ecological Modelling}, \emph{168}, 145--152.

\bibitem{WilhelmHollunder2007}
Wilhelm, T. \& Hollunder, J. (2007).
\newblock Information theoretic description of networks.
\newblock \emph{Physica A}, \emph{385}, 385--396.

\bibitem{Wolfetal2000}
Wolf, J., Brodie, E., \& Wade, M. (Eds.)  (2000).
\newblock \emph{Epistasis and the Evolutionary Process}.
\newblock Oxford: Oxford University Press.

\end{thebibliography}
%\newpage
%\section*{Figures}
%Fig. 1

%Fig. 2

%Fig. 3

%Fig. 4

%Fig. 5

%Fig. 6

%Fig. 7
%Fig. 8

%\newpage

%\section*{Tables}
%Tab. 1
% Requires the booktabs if the memoir class is not being used

\end{document}